\documentclass[preprint,aps,12pt,showpacs,nofootinbib,tightenlines,amsmath,amssymb]{revtex4}

\usepackage{multirow}
\usepackage{epsfig}

\begin{document}


\newcommand{\CDP}{\hat{\cal D}\hspace{-0.27cm}\slash_v}
\newcommand{\CDPL}{\overleftarrow{\hat{\cal D}}\hspace{-0.29cm}\slash_v}
\newcommand{\CDPN}{{\cal D}\hspace{-0.26cm}\slash_v}
\newcommand{\CDPNL}{\overleftarrow{\cal D}\hspace{-0.29cm}\slash_v}
\newcommand{\Dbot}{{/\!\!\!\!D}\hspace{-3pt}_{\bot}}
\newcommand{\DC}{D_{\! \bot}}
\newcommand{\DS}{D\hspace{-0.25cm}\slash}
\newcommand{\DSP}{\not\!{v} v\cdot D}
\newcommand{\dsp}{D\hspace{-0.25cm}\slash_{\|}}
\newcommand{\DSC}{D\hspace{-0.25cm}\slash_{\bot}}
\newcommand{\DSCX}{D\hspace{-0.26cm}\slash_{\bot}}
\newcommand{\DSCXL}{\overleftarrow{D}\hspace{-0.29cm}\slash_{\bot}}
\newcommand{\DSPX}{\not\!{v} v\cdot D}
\newcommand{\Dslashbot}{{/\!\!\!\!D}\hspace{-3pt}_{\bot}}
\newcommand{\Dslash}{/\!\!\!\!D}
\def\fvcb{${\cal F}(1)|V_{\rm cb}|$}
\newcommand{\hm}{\hat{m}_b}
\newcommand{\kslash}{\not\!{k}}
\newcommand{\kslashbot}{{/\!\!\!\!k}\hspace{-3pt}_{\bot}}
\newcommand{\mbh}{\hat{m}_b}
\newcommand{\MQ}{m_Q}
\newcommand{\MQP}{m_{Q^{\prime}}}
\newcommand{\non}{\nonumber \\}
\def\Ova{O^q_{V-A}}
\def\Osp{O^q_{S-P}}
\newcommand{\oDbot}{\overleftarrow{/\!\!\!\!D}\hspace{-5pt}_{\bot}}
\newcommand{\oDSP}{\not\!{v} v\cdot \overleftarrow{D}}
\newcommand{\odsp}{\overleftarrow{D}\hspace{-0.29cm}\slash_{\|}}
\newcommand{\oDslashbot}{\overleftarrow{/\!\!\!\!D}\hspace{-5pt}_{\bot}}
\newcommand{\omiga}{\omega}
\newcommand{\pbot}{\partial\hspace{-3pt}_\bot}
\newcommand{\PP}{{1 + v\hspace{-0.2cm}\slash \over 2}}
\newcommand{\PM}{{1 - v\hspace{-0.2cm}\slash \over 2}}
\newcommand{\PSC}{\partial\hspace{-0.2cm}\slash_{\bot}}
\newcommand{\PSP}{\partial\hspace{-0.2cm}\slash_{\|}}
\newcommand{\ptslashbot}{\partial\hspace{--0.2cm}\slash_\bot}
\newcommand{\QVBP}{\bar{Q}^{+}_{v^{\prime}} }
\newcommand{\QV}{Q^{+}_v}
\newcommand{\QVB}{\bar{Q}^{+}_v}
\newcommand{\QVH}{\hat{Q}_v}
\newcommand{\QVHB}{\bar{\hat{Q}}_v}
\newcommand{\QVHF}{\hat{Q}^{-}_v}
\newcommand{\QVHMP}{\hat{Q}^{(\mp)}_v}
\newcommand{\QVHFB}{\bar{\hat{Q}}_v{\vspace{-0.3cm}\hspace{-0.2cm}{^{(-)}} }}
\newcommand{\QVHPM}{\hat{Q}^{(\pm)}_v}
\newcommand{\QVPHFB}{\bar{\hat{Q}}_{v^{\prime}}{\vspace{-0.3cm}\hspace{-0.2cm}{^{\prime (-)}} } }
\newcommand{\QVPHZB}{\bar{\hat{Q}}_{v^{\prime}}{\vspace{-0.3cm}\hspace{-0.2cm}{^{\prime (+)}}} }
\newcommand{\QVHMPB}{\bar{\hat{Q}}_v{\vspace{-0.3cm}\hspace{-0.2cm}{^{(\mp)}} }  }
\newcommand{\QVHPMB}{\bar{\hat{Q}}_v{\vspace{-0.3cm}\hspace{-0.2cm}{^{(\pm)}} }}
\newcommand{\QVHZB}{\bar{\hat{Q}}_v{\vspace{-0.3cm}\hspace{-0.2cm}{^{(+)}} } }
\newcommand{\QVHZ}{\hat{Q}^{+}_v}
\newcommand{\qslash}{\not\!{q}}
\newcommand{\uv}{/\!\!\!\!\hspace{1pt}v}
\newcommand{\uvslash}{/\!\!\!\!\hspace{1pt}v}
\newcommand{\VS}{\not\!{v}}
\newcommand{\vslash}{\not\!{v}}


\draft
\title{ $B\to D'_0 (D'_1)\ell \bar{\nu}$ decays in HQEFT }
\author{ W.Y. Wang\footnote{E-mail address: wangwenyu@tsinghua.org.cn} }
\address{ Department of Physics, University of Science and Technology Beijing, Beijing 100083, China \\
Kavli Institute for Theoretical Physics China, CAS, Beijing 100190, China }

\begin{abstract}

Semileptonic B decays into excited charmed mesons $D'_0$ and
$D'_1$ are studied in the framework of heavy quark effective field
theory (HQEFT) up to order $1/m_Q$. They are characterized by a
single leading Isgur-Wise function $\tau$ and several wave
functions arising at $1/m_Q$ order. $\tau$ and the
$1/m_Q$ order functions $\chi^b_0$, $\chi^c_0$ related to the
kinetic energy operators are evaluated through QCD sum rule approach;
zero recoil values of the $1/m_Q$ order functions $\kappa_1$, $\kappa_2$, $\kappa'_1$ and $\kappa'_2$
are extracted from the meson
masses; and the branching ratios are found to be suppressed by the
$1/m_Q$ corrections. It is concluded that the next leading order
wave functions can be significant.
However it does not change the previous prediction that the
production rate of $j_l^P=\frac{3}{2}^+$ charmed mesons dominates
over that of $\frac{1}{2}^+$ doublets.

\end{abstract}

\vspace{0.5cm}


\maketitle


\section{Introduction}

Semileptonic B decays are important in extracting the CKM matrix
elements and exploring CP violation. Presently the most promising
approach to determine $|V_{cb}|$ is to study either the inclusive
semileptonic B decays or the exclusive decays to the ground state
charmed mesons, $B\to D(D^*)\ell\bar{\nu}$. However the precision
of these study depends on both the experimental measurements and
the theoretical methods probing the nonperturbative effects of
strong interaction. Semileptonic B decays into excited charmed
mesons contain the main background for measuring the decays into
$D$ and $D^*$, and they are important in relating the inclusive B
decays to the sum of exclusive channels. To get precision
knowledge on B physics it needs to study the decays into excited
mesons from both experimental and theoretical aspects.

For a hadron containing a single heavy quark Q (b or c) and any
number of light quarks (u, d, s), the heavy quark spin $s_Q$ is
decoupled from the total angular momentum of the light degrees of
freedom $j_l$ in the heavy quark limit $m_Q\to \infty$. So $j_l$
becomes a good quantum number in this limit. Consequently, charmed
mesons are usually classified by $j_l$ and parity. The ground
state pseudoscalar and vector mesons ($D$, $D^*$) have
$j_l^P=\frac{1}{2}^-$. $D_1$ and $D^*_2$ belong to the
$j_l^P=\frac{3}{2}^+$ doublet, while $D'_0$ (or written as $D^*_0$
in some references) and $D'_1$ are the $\frac{1}{2}^+$ one.

In the recent years experiments made rapid progress on charmed
meson spectroscopy, especially for the four lightest excited
charmed mesons ($D_1$, $D^*_2$, $D'_0$ and $D'_1$) as well as
their counterparts of $c\bar{s}$ states. The $\frac{3}{2}^+$
doublet mesons have narrow widths and their masses are known
precisely: $m_{D_1}=2420\mbox{MeV}$ and $m_{D^*_2}=2460\mbox{MeV}$
\cite{PDG}. Broad charmed resonances are observed in $D\pi $ and
$D^*\pi$ systems by BELLE \cite{BELLE}, FOCUS \cite{FOCUS} and
CLEO \cite{CLEO} Collaborations. The masses and widths of the
$\frac{1}{2}^+$ doublets can be obtained from these measurements,
nevertheless they still suffer from large
uncertainties\cite{PCola0407137}. The branching ratios for
semileptonic decays $B\to D_1(D^*_2)\ell \bar{\nu}$ are reported
by CLEO \cite{CLEO2}, ALEPH \cite{ALEPH}, D0 \cite{VMAD0} and
BELLE \cite{BELLE091503} Collaborations. Though not being
confirmed, $B\to D'_0\ell \bar{\nu}$ decay ratio is obtained
recently by BELLE \cite{BELLE091503}. On the other hand these
decays are studied by theorists via different approaches, among
which are the operator product expansion (OPE)
\cite{NUra501,IIB975}, Lattice QCD \cite{DBec609} and quark models
\cite{FJug094010,AKLeib9703213,AKLeib}.
Note the $1/m_Q$ order corrections in the usual heavy quark expansion
has been considered in the early work \cite{AKLeib9703213,AKLeib}.
QCD sum rule method is also applied to
calculate the form factors. Refs.\cite{PCola116005,YBDai9807461}
studied the semileptonic B decays into excited charmed mesons at
the leading order of heavy quark expansion (HQE), and the $1/m_Q$
order contributions for $B\to D_1(D^*_2)\ell \bar{\nu}$ are
considered in Refs.\cite{MQHuang0102299} and \cite{WYWang2505}
using different framework of HQE.

Generally, the theoretical calculations in the $m_Q \to \infty$
limit predict that the production of $\frac{3}{2}^+$ doublets
dominates over that of $\frac{1}{2}^+$ doublets in semileptonic B
decays. As illustrated in Ref.\cite{IIB975}, for reasonable values
of the Isgur-Wise function, the rate $\Gamma(B\to D'_0(D'_1)\ell
\bar{\nu})$ falls far below $\Gamma(B\to D_1(D^*_2)\ell
\bar{\nu})$. However, BELLE indicates in Ref.\cite{BELLE091503} a
large branching ratio for B decay to the wide $D'_0$ state. If
this result is confirmed, the previous throries or models need to
be improved or corrected to explain it.
Generally speaking, the predictions derived in the $m_Q \to \infty$ limit should always
be supported by the estimation of $1/m_Q$ corrections, which turn out to be sizable in some specific situations. For example, it is known that the leptonic decay constants of heavy mesons receive
considerable $1/m_Q$ corrections \cite{neubprd1076,JMFlynn402}.
Calculations in different approaches also indicate
large $1/m_Q$ corrections to the $B\to D_1 \ell \bar{\nu}$ decay rate
\cite{AKLeib,MQHuang0102299,WYWang2505}.
In the case of $B\to D'_0(D'_1)\ell \bar{\nu}$ transitions, one may
ask whether the great enhancement of the production rate for
$\frac{1}{2}^+$ states is due to the finite mass corrections in the
HQE.

In this paper, the semileptonic B decays into the $\frac{1}{2}^+$
charmed meson doublet ($D'_0$, $D'_1$) are studied in the
framework of HQEFT\cite{YLWu,WYWangweak,WYWangconsi,YLWunew} that
performs a complete decomposition of quantum fields and therefore
includes the heavy quark-antiquark coupling effects in the finite
mass corrections. In Sec.\ref{formulation} we present the
formulation of HQE to the decays $B\to D'_0(D'_1)\ell \bar{\nu}$.
Up to the order of $1/m_Q$, the relevant form factors are given by
universal wave functions that are heavy flavor independent. In
Sec.\ref{sumrule} QCD sum rule approach is used to evaluate the
leading Isgur-Wise function $\xi$ and the next leading order wave
functions $\chi^b_0$, $\chi^c_0$ relevant to the kinetic energy
operator. The QCD sum rule for leptonic decay constant of
$\frac{1}{2}^+$ mesons is also derived. Sec.\ref{analysis}
analyzes the sum rules and gives numerical results. Finally a brief
summary is given in Sec.\ref{summary}.

\section{$B\to D'_0(D'_1) \ell \nu$ wave functions in HQEFT}\label{formulation}

The weak matrix elements relevant to $B\to D'_0(D'_1)\ell
\bar{\nu}$ decays can be characterized by form factors as
\begin{eqnarray}
\label{defformfactor}
 \langle D'_0(v')|\bar{c}\gamma^\mu b|B(v) \rangle &=&0 , \nonumber \\
 \langle D'_0(v')|\bar{c} \gamma^\mu
\gamma^5 b|B(v) \rangle &=&\sqrt{m_{D'_0} m_B}
   (g_+ (v^\mu +v'^\mu)+g_- (v^\mu-v'^\mu) ), \nonumber\\
\langle D'_1(v',\epsilon^*)|\bar{c}\gamma^\mu b|B(v) \rangle
&=&\sqrt{m_{D'_1} m_B}
  g_{V_1} \epsilon^{*\mu}+(g_{V_2} v^\mu + g_{V_3} v'^\mu )
  (\epsilon^* \cdot v) , \nonumber\\
 \langle D'_1(v',\epsilon^*)|\bar{c} \gamma^\mu
\gamma^5 b|B(v) \rangle &=& \sqrt{m_{D'_1} m_B}
    i g_A \epsilon^{\mu\alpha\beta\gamma}\epsilon^*_{\alpha}
   v_\beta v'_\gamma .
\end{eqnarray}
The initial and final states $B$ and $D'_{0(1)}$ are treated as
heavy hadrons with the momentum $m_Bv$ and $m_{D'_{0(1)}} v'$,
respectively. The form factors $g_i$ are dimensionless functions
of the product of velocities, $y=v\cdot v'$, and $\epsilon^*$ is
the polarization vector of the axial vector meson $D'_1$. The
differential decay rates are given by
\begin{eqnarray}
\frac{d\Gamma(B\to D'_0 \ell \bar\nu)}{dy}&=& \frac{G_F^2
|V_{cb}|^2 m^5_B }{48 \pi^3} {r'_0}^3
   (y^2-1)^{3/2} [ (1+r'_0 ) g_+ -(1-r'_0) g_- ]^2,  \\
\frac{d\Gamma(B\to D'_1 \ell \bar\nu)}{dy}&=& \frac{G^2_F
|V_{cb}|^2 m^5_B }{48 \pi^3} {r'_1}^{3}
   (y^2-1)^{1/2} \{ 2(1-2 r'_1 y+{r'_1}^2 ) [g^2_{V_1}+(y^2-1)
   g^2_A]\nonumber \\
&&  +[(y-r'_1) g_{V_1}+(y^2-1) (g_{V_3}+r'_1 g_{V_2}) ]^2 \}
\end{eqnarray}
with $r'_0=\frac{m_{D'_0}}{m_B}$ and $r'_1=\frac{m_{D'_1}}{m_B}$.

In the framework of HQEFT the matrix elements in QCD can be
expanded in powers of $1/m_Q$. Generally, the HQE of the matrix
elements responsible for heavy meson leptonic decays and for
transitions between heavy mesons can be written as
\cite{WYWangweak,WYWangconsi}
\begin{eqnarray}
 \label{eq:matrixexp1}
 &&\sqrt{\frac{\bar{\Lambda}_M}{m_M}}  \langle 0|\bar{q}\Gamma Q|M \rangle  \to
   \langle 0|\bar{q} \Gamma \QV|M_v \rangle -\frac{1}{2m_Q}
    \langle 0|\bar{q}\Gamma\frac{1}{i \dsp}(i\DSC)^2\QV |M_v \rangle
  +O(1/m^2_Q),  \\
\label{eq:matrixexp2}
 && \sqrt{\frac{\bar{\Lambda}_{M'} \bar{\Lambda}_{M} }{m_{M'} m_M}}
   \langle M'|\bar{Q}' \Gamma Q|M \rangle  \to
  \langle M'_{v'}|\QVBP \Gamma \QV|M_v \rangle -\frac{1}{2m_Q}
   \langle M'_{v'}|\QVBP\Gamma\frac{1}{i \dsp}(i\DSC)^2\QV |M_v \rangle
 \nonumber \\
&&\hspace{2cm}  -\frac{1}{2m_{Q'}}  \langle M'_{v'}|\QVBP
  (-i \! \! \stackrel{\hspace{-0.1cm}\leftarrow}{\DSC})^2
   \frac{1}{-i \!\! \stackrel{\hspace{-0.1cm}\leftarrow}{\dsp}}
   \Gamma \QV|M_v \rangle  +O(1/m^2_Q)
 \end{eqnarray}
with the definition
\begin{eqnarray}
&& D^\mu_{\! \|} = v^\mu v\cdot D , \\
&& D^\mu_{\! \bot} = D^\mu -v^\mu v\cdot D , \\
&& \int \kappa \overleftarrow{D} \mbox{}^\mu \varphi = -\int
\kappa D^\mu \varphi .
\end{eqnarray}
$M(M')$ can be any ground or excited heavy meson containing a
single heavy quark. $Q$ is the field in the full QCD Lagrangian,
and $Q^{+}_v$ is the effective heavy quark field in HQEFT,
carrying only the residual momentum $k=p_Q-m_Q v$. $| M \rangle$
is the meson state in the full theory, while $|M_v \rangle$ is an
effective state defined in the HQEFT so as to display the heavy
quark spin-flavor symmetry. They are normalized as
\begin{eqnarray}
\label{statenor1}
&&\langle M|\bar{Q}\gamma^\mu Q|M\rangle = 2m_M
v^\mu , \\
\label{statenor2}
 && \langle M_v|\bar{Q}^{+}_{v} \gamma^\mu
Q^{+}_{v}|M_v\rangle = 2\bar{\Lambda} v^\mu  ,
\end{eqnarray}
where $\bar{\Lambda}_M \equiv m_M-m_Q$ is the mass difference
between the heavy meson and heavy quark, and
\begin{equation}
\label{eq:bindrelat} \bar{\Lambda} = \lim_{m_{Q}\to \infty}
\bar{\Lambda}_H = \lim_{m_{Q}\to \infty} (m_M-m_Q)
\end{equation}
is the heavy flavor independent binding energy. The state $| M_v
\rangle$ defined in this way is irrelevant to the heavy quark mass
and related to $|M \rangle$ via
\begin{equation}
    \label{eq:statedef}
     \frac{1}{\sqrt{m_{M^{\prime}}m_{M}}} \langle M^{\prime}|\bar{Q}^{\prime} \Gamma Q | M \rangle =
 \frac{1}{\sqrt{\bar{\Lambda}_{M^{\prime}} \bar{\Lambda}_M}} \langle
 M^{\prime}_{v^{\prime}}|
          J_{eff} e^{i\int d^4x {\cal L}_{eff}} | M_v \rangle ,
\end{equation}
where ${\cal L}_{eff}$ is the HQEFT Lagrangian and $J_{eff}$ is
the effective current for $\bar{Q}\Gamma Q$, generally also
written as an expansion in $1/m_Q$. For more details of HQEFT we
refer to Refs.\cite{YLWu,WYWangweak,WYWangconsi,YLWunew}.

Due to the heavy quark symmetry, form factors for heavy-to-heavy
transition matrix elements can be parameterized by a set of wave
functions, which are universal in that they are heavy flavor and
spin independent. To define these wave functions one may use the
following spin wave functions for the ground ($\frac{1}{2}^-$) and
excited ($\frac{1}{2}^+$) states:
\begin{eqnarray}
     \label{eq:spinwave1}
       {\cal M}_v=\sqrt{\bar{\Lambda}}P_{+}
         \left\{
           \begin{array}{cl}
              -\gamma^{5}, & \mbox{for $B$} ,\; \; \\
              \epsilon\hspace{-0.15cm}\slash, & \mbox{for $B^*$}, \; \;
           \end{array}
         \right.
     \end{eqnarray}
 \begin{eqnarray}
   \label{eq:spinwave3}
   {\cal K}_v=\sqrt{\bar{\Lambda}'}P_{+}
    \left\{
       \begin{array}{cl}
       1, & \mbox{for $D'_0$} ,\; \; \\
      -\epsilon\hspace{-0.15cm}\slash \gamma^5, & \mbox{for $D'_1$} ,\; \;
           \end{array}
         \right.
     \end{eqnarray}
where $\bar{\Lambda}=\bar{\Lambda}_{\frac{1}{2}^-}$ and
$\bar{\Lambda}'=\bar{\Lambda}_{\frac{1}{2}^+}$ are the binding
energies of the $\frac{1}{2}^-$ and $\frac{1}{2}^+$ doublets,
respectively, and $P^{(\prime)}_{\pm} \equiv (1\pm
\VS^{(\prime)})/2$. Then in HQEFT the matrix elements between
$\frac{1}{2}^+$ states can be parameterized as
\begin{eqnarray}
\label{parnor}
 \langle K_{v'}|\QVBP \gamma^\mu \QV|{ K}_v \rangle &=& \xi'(y) Tr[\bar{\cal{K}}_{v'} \gamma^\mu
    {\cal{K}}_{v} ], \nonumber\\
 \langle  K_{v'}|\QVBP \gamma^\mu \frac{P_+}{i v\cdot D} \DC^2 \QV|
  K_v \rangle &=& -\kappa'_1(y)
    \frac{1}{\bar{\Lambda}'}
    Tr[\bar{\cal K}_{v'} \gamma^\mu {\cal{K}}_{v} ] ,\nonumber\\
 \langle { K}_{v'}|\QVBP \gamma^\mu \frac{P_+}{i v\cdot D} \frac{i}{2} \sigma_{\alpha \beta} F^{\alpha\beta}
   \QV|{K}_v \rangle &=& \frac{1}{\bar{\Lambda}'}
  Tr[ \kappa'^{\alpha\beta}(v,v') \bar{{\cal{K}}}_{v'} \gamma^\mu P_+
    \frac{i}{2} \sigma_{\alpha\beta} {\cal{K}}_{v} ]
\end{eqnarray}
with $\bar{{\cal{K}}}_{v'}\equiv \gamma^0 {\cal{K}}^{\dagger}_{v'}
\gamma^0$. The states $K_v$ and $K_{v'}$ can be either of the two
mesons belonging to the $\frac{1}{2}^+$ doublet. The tensor
$\kappa'^{\alpha\beta}(v,v')$ are decomposed as
\begin{eqnarray}
\label{decom} \kappa'^{\alpha\beta}(v,v')&=&i \kappa'_2
  \sigma^{\alpha\beta}-\kappa'_3 (\gamma^\alpha v'^{\beta}
  -\gamma^\beta v'^{\alpha}) +\kappa'_4 (\gamma^\alpha v^\beta
  -\gamma^\beta v^\alpha)+\kappa'_5 (v^\alpha v'^\beta
 -v^\beta v'^\alpha)
\end{eqnarray}
with $\xi'$ and $\kappa'_i$ being scalar functions of $y=v\cdot
v'$. Carrying out the trace calculation in (\ref{parnor}) and
setting $v'=v$, one gets from Eqs.(\ref{eq:matrixexp2}),
(\ref{statenor1}), (\ref{statenor2}), (\ref{eq:statedef}) and
(\ref{parnor})
\begin{eqnarray}
 2m_{D'_0} v^\mu&=&\frac{m_{D'_0}}{\bar{\Lambda}_{D'_0}}
  \{2\bar{\Lambda}' \xi'(1)-\frac{2}{m_c}
  (\kappa'_1(1)+3\kappa'_2(1)) \} v^\mu  , \\
2m_{D'_1} v^\mu&=&\frac{m_{D'_1}}{\bar{\Lambda}_{D'_1}}
  \{-2\bar{\Lambda}' \xi'(1) +\frac{2}{m_c}
  (\kappa'_1(1) - \kappa'_2(1))  \} (\epsilon^* \cdot \epsilon
  )v^\mu .
\end{eqnarray}
Since the Isgur-Wise function $\xi'(1)$ satisfies the
normalization condition $ \xi'(1)=1  $, the above equations yield
\begin{eqnarray}
\label{binding1} \bar{\Lambda}_{D'_0} &=& \bar{\Lambda}'
-\frac{1}{m_c}
   (\kappa'_1(1)+3\kappa'_2(1)), \\
\label{binding2} \bar{\Lambda}_{D'_1} &=& \bar{\Lambda}'
-\frac{1}{m_c}
   (\kappa'_1(1)-\kappa'_2(1)) ,
\end{eqnarray}
which are quite similar to those relations for the
$j_l^P=\frac{1}{2}^-$ ground state bottom mesons
\cite{WYWangweak,WYWangconsi}:
\begin{eqnarray}
\label{binding3} \bar{\Lambda}_{B} &=& \bar{\Lambda}
-\frac{1}{m_b}
   (\kappa_1(1)+3\kappa_2(1)) , \\
\label{binding4} \bar{\Lambda}_{B^*} &=& \bar{\Lambda}
-\frac{1}{m_b}
   (\kappa_1(1)-\kappa_2(1)) .
\end{eqnarray}

For $B\to D'_0(D'_1)$ transitions, the relevant matrix elements
can be parameterized in HQEFT as
\begin{eqnarray}
\label{wavefunctiondef}
 \langle K_{v'}|\QVBP \Gamma \QV|B_v \rangle &=&\tau(y)
  Tr[\bar{\cal K}_{v'} \Gamma {\cal M}_v ]  ,\nonumber \\
\langle K_{v'}|\QVBP \Gamma \frac{P_+}{i v\cdot D } \DC^2 \QV|B_v \rangle
 &=&-\chi^{b}_0(y) \frac{1}{\bar{\Lambda}}
 Tr[ \bar{{\cal{K}}}_{v'} \Gamma {\cal{M}}_v ] , \nonumber \\
 \langle K_{v'}|\QVBP
 {\stackrel{\hspace{-0.1cm}\leftarrow}D_{\! \bot}} \mbox{}^{\!\!\! 2}
   \frac{P'_+}{-i v' \cdot \stackrel{\hspace{-0.1cm}\leftarrow}D }
\Gamma \QV|B_v \rangle
  &=&-\chi^{c}_0(y) \frac{1}{\bar{\Lambda}'}
 Tr[\bar{{\cal{K}}}_{v'} \Gamma {\cal{M}}_v ] , \nonumber \\
 \langle K_{v'}|\QVBP \Gamma \frac{P_+}{i v\cdot D } \frac{i}{2} \sigma_{\alpha\beta} F^{\alpha\beta}
   \QV|B_v \rangle &=& -\frac{1}{\bar{\Lambda}} Tr[ R^{b}_{\alpha\beta}
  (v,v') \bar{\cal{K}}_{v'} \Gamma P_+ i\sigma_{\alpha\beta}
   {\cal{M}}_v ] ,\nonumber \\
 \langle K_{v'}|\QVBP \frac{i}{2}\sigma_{\alpha\beta} F^{\alpha\beta}
   \frac{P'_+}{-i v'\cdot \stackrel{\hspace{-0.1cm}\leftarrow} D}
\Gamma \QV|B_v \rangle
 &=&-\frac{1}{\bar{\Lambda}'} Tr[ R^{c}_{\alpha\beta} (v,v')
  \bar{\cal{K}}_{v'} i\sigma^{\alpha\beta} P'_+ \Gamma {\cal{M}}_v ] ,
\end{eqnarray}
where the Lorentz tensors $R^{b(c)}_{\alpha\beta}(v,v')$ can be
decomposed as
\begin{eqnarray}
R^{b}_{\alpha\beta}(v,v')&=&\chi^{b}_1 \gamma_\alpha \gamma_\beta
 +\chi^{b}_2 v'_\alpha \gamma_\beta ,  \\
R^{c}_{\alpha\beta}(v,v')&=&\chi^{c}_1 \gamma_\alpha \gamma_\beta
  +\chi^{c}_2 v_\alpha \gamma_\beta .
\end{eqnarray}
The wave functions $\tau$ and $\chi^{b(c)}_i
\hspace{0.2cm}(i=0,1,2)$ depend on y. $\tau$ is dimensionless,
while $\chi^{b(c)}_i$ has mass dimension two.

HQE for the form factors $g_i$ is then obtained from
(\ref{defformfactor}), (\ref{eq:matrixexp2}) and
(\ref{wavefunctiondef}). Up to the order of $1/m_Q$ one has
\begin{eqnarray}
\label{formfactorHQE}
g_+ &=&0 , \nonumber\\
g_- &=& \tilde{\tau}+\frac{\tau}{2m_b\bar{\Lambda} }
  (\kappa_1(1)+3\kappa_2(1))+\frac{\tau}{2m_c \bar{\Lambda}'}
  (\kappa'_1(1)+3\kappa'_2(1))
  -\frac{1}{m_b\bar{\Lambda}} \chi^{b}  \nonumber \\
  && -\frac{1}{m_c \bar{\Lambda}'} [3 \chi^{c}_1- \chi^{c}_2
  (1+y) ] , \nonumber \\
g_{V_1}&=&[\tilde{\tau}+\frac{\tau}{2m_b\bar{\Lambda} }
  (\kappa_1(1)+3\kappa_2(1))+\frac{\tau}{2m_c \bar{\Lambda}'}
  (\kappa'_1(1)-\kappa'_2(1)) -\frac{1}{m_b\bar{\Lambda}} \chi^{b} \nonumber \\
  && +\frac{1}{m_c \bar{\Lambda}'} \chi^{c}_1
  ] (y-1) , \nonumber \\
g_{V_2}&=& \frac{\chi^{c}_2}{m_c \bar{\Lambda}'}, \nonumber \\
g_{V_3}&=&- \tilde{\tau} -\frac{\tau}{2m_b\bar{\Lambda} }
  (\kappa_1(1)+3\kappa_2(1)) -\frac{\tau}{2m_c \bar{\Lambda}'}
  (\kappa'_1(1)-\kappa'_2(1))
  +\frac{1}{m_b\bar{\Lambda}} \chi^{b}  \nonumber \\
  && -\frac{1}{m_c \bar{\Lambda}'} (\chi^{c}_1-\chi^{c}_2), \nonumber \\
g_{A}&=& \tilde{\tau}+\frac{\tau}{2m_b\bar{\Lambda} }
  (\kappa_1(1)+3\kappa_2(1))+\frac{\tau}{2m_c \bar{\Lambda}'}
  (\kappa'_1(1)-\kappa'_2(1))
  -\frac{1}{m_b\bar{\Lambda}} \chi^{b}  \nonumber \\
  && +\frac{1}{m_c \bar{\Lambda}'} \chi^{c}_1
\end{eqnarray}
with
\begin{eqnarray}
\label{taucor}
 \tilde{\tau}&=&\tau-\frac{\chi^{b}_0}{2m_b \bar{\Lambda}}
  -\frac{\chi^{c}_0}{2m_c \bar{\Lambda}'},  \\
 \chi^{b}&=&3\chi^{b}_1-(1+y) \chi^{b}_2 .
\end{eqnarray}
Here $\kappa^{(\prime)}_i(1) \;  (i=1,2)$ are the zero recoil
values of $\kappa^{(\prime)}_i$, whereas other wave functions and
form factors depend on the variable $y=v\cdot v'$.

\section{QCD Sum Rules for Wave Functions}\label{sumrule}

As can be seen in (\ref{formfactorHQE}), in the heavy quark limit
all form factors simply reduce to the Isgur-Wise function $\tau$.
Among the 6 functions $\chi^{b(c)}_{i}(i=0,1,2)$ of order $1/m_Q$,
$\chi^{b(c)}_{1(2)}$ are defined in (\ref{wavefunctiondef}) by the
chromomagnetic operators. Contributions from such operators are
generally expected to be very small, which is supported by the
relativistic quark model \cite{DEbert} and QCD sum rule study
\cite{MNeubZY}. Here we mainly focus on the functions $\chi^{b}_0$
and $\chi^{c}_0$, which are defined by the matrix elements of the
kinetic energy operators. Since the kinetic operators preserve
heavy quark spin symmetry, $\chi^{b(c)}_{0}$ simply correct the
leading Isgur-Wise function $\tau$ in the way of
Eq.(\ref{taucor}).

In order to calculate $\tau$ and $\chi^{b}_0$, $\chi^{c}_0$, we
study the following three-point correlation functions
\begin{eqnarray}
\label{correlator1}
 \Xi^\tau &=&i^2 \int d^4x d^4z e^{i (k'\cdot x-k\cdot z)}  \langle 0|
  T\{ J_{0,+,1/2} (x),
  (\bar{Q}^{+}_{v'}\Gamma Q^{+}_{v})(0), J^{\dagger}_{0,-,1/2}(z) \}|0 \rangle , \\
\label{correlator3} \Xi^{\chi^b_0}  &=& i^2 \int d^4x d^4z e^{i
(k'\cdot x-k\cdot z)}  \langle 0|T\{ J_{0,+,1/2}(x),
  (\bar{Q}^{+}_{v'}\Gamma \frac{P_+}{iv\cdot D} \DC^2 Q^{+}_{v})(0), J^{\dagger}_{0,-,1/2}(z) \}|0 \rangle, \\
\label{correlator5} \Xi^{\chi^c_0} &=& i^2 \int d^4x d^4z e^{i
(k'\cdot x-k\cdot z)}  \langle 0|T\{ J_{0,+,1/2}(x),
  (\bar{Q}^{+}_{v'}
  \stackrel{\hspace{-0.1cm}\leftarrow}D_{\!\bot}\mbox{}^{\!\! \! \! \! 2}
  \frac{P'_{+}}
  {-iv'\cdot \stackrel{\hspace{-0.2cm}\leftarrow} D }\Gamma Q^{+}_{v})(0),
  J^{\dagger}_{0,-,1/2}(z) \}|0 \rangle,
\end{eqnarray}
where $k$ and $k'$ are the residual momenta of the heavy quarks.
$\Gamma$ should be $\gamma^\mu$ and $\gamma^\mu \gamma^5$ for
vector and axial vector heavy quark currents respectively.
$J_{j,P,j_l}$ with $j$ the total spin of the meson should be
proper interpolating currents for the heavy-light mesons. One set
of such currents are proposed in Ref.\cite{YBDai9609436}. One can
use
\begin{eqnarray}
\label{currentf0} J^{\dagger}_{0,-,1/2}&=&\sqrt{\frac{1}{2}}
\bar{Q}^{+}_{v} \gamma^5
q ,  \\
\label{currentf1}
J^{\dagger}_{1,-,1/2}&=&\sqrt{\frac{1}{2}}\bar{Q}^{+}_{v}
\gamma^\alpha_\bot q ,
\end{eqnarray}
for the $\frac{1}{2}^-$ ground state doublet, and
\begin{eqnarray}
\label{currentz0x} &&J^{\dagger}_{0,+,1/2}=\sqrt{\frac{1}{2}}\QVBP
q , \\
\label{currentz1x}
 &&J^{\dagger}_{1,+,1/2}=\sqrt{\frac{1}{2}}\QVBP
\gamma^5
  \gamma^\alpha_\bot q
\end{eqnarray}
or
\begin{eqnarray}
\label{currentz0} &&J^{\dagger}_{0,+,1/2}=\sqrt{\frac{1}{2}}\QVBP
(-i)\DSC q , \\
\label{currentz1} &&J^{\dagger}_{1,+,1/2}=\sqrt{\frac{1}{2}}\QVBP
\gamma^5
  \gamma^\alpha_\bot (-i) \DSC q
\end{eqnarray}
for the $\frac{1}{2}^+$ doublet. $\gamma^\alpha_\bot$ is defined
as $\gamma^\alpha_{\bot} = \gamma^\alpha - v^\alpha \VS$. In
Eqs.(\ref{correlator1})-(\ref{correlator5}) $J_{0,+,1/2}$ is used
in the three-point functions. Of course one can substitute
$J_{1,+,1/2}$ for $J_{0,+,1/2}$ in the evaluation, which does not
make difference to the results for wave functions $\tau$ and
$\chi^{b(c)}_0$, as required by the heavy quark symmetry.

The formulae in (\ref{correlator1})-(\ref{correlator5}) are
analytic functions of the variables $\omega=2 v \cdot k$ and
$\omega'=2 v' \cdot k'$ with discontinuities for their positive
values. The phenomenological representation for these correlators
can be obtained by inserting the complete set of intermediate
states with the same quantum numbers as the currents $J_{0,+,1/2}$
and $J_{0,-,1/2}$. Isolating the pole terms of the lowest states
we get
\begin{eqnarray}
\label{phentau} \Xi^{\tau}_{phen} &=&
   \frac{ \langle 0|J_{0,+,1/2}|K_{v'} \rangle
   \langle K_{v'}|\bar{Q}^{+}_{v'} \Gamma \QV|B_v\rangle \langle B_v|J^{\dagger}_{0,-,1/2} |0\rangle }
   {(2\bar{\Lambda}-\omega-i\epsilon)(2\bar{\Lambda}'-\omega'-i\epsilon) \bar{\Lambda}\bar{\Lambda}'}
   + \mbox{higher resonances} ,\\
\label{phenb} \Xi^{\chi^b_0}_{phen} &=&
   \frac{ \langle 0|J_{0,+,1/2}|K_{v'} \rangle
   \langle K_{v'}|\bar{Q}^{+}_{v'} \Gamma\frac{P_{+}}{iv\cdot D}\DC^2 \QV|B_v\rangle \langle B_v|J^{\dagger}_{0,-,1/2} |0\rangle }
   {(2\bar{\Lambda}-\omega-i\epsilon)(2\bar{\Lambda}'-\omega'-i\epsilon)\bar{\Lambda}\bar{\Lambda}'}
   + \mbox{higher resonances} ,\\
\label{phenc} \Xi^{\chi^c_0}_{phen} &=&
  \frac{ \langle 0|J_{0,+,1/2}|K_{v'} \rangle
  \langle K_{v'}| \bar{Q}^{+}_{v'}
  \stackrel{\hspace{-0.1cm}\leftarrow}{D_{\!\bot}} \mbox{}^{\!\!\!\!\! 2}
  \frac{P'_+}
  {-iv'\cdot \stackrel{\hspace{-0.2cm}\leftarrow} D}\Gamma \QV |B_v \rangle \langle B_v|J^{\dagger}_{0,-,1/2} |0\rangle }
 {(2\bar{\Lambda}-\omega-i\epsilon)(2\bar{\Lambda}'-\omega'-i\epsilon)\bar{\Lambda}\bar{\Lambda}'}
  +\mbox{higher resonances},
\end{eqnarray}
where the first term in each equation is a double-pole
contribution, and the second term takes into account higher states
and continuum contributions. Using $\Gamma=\gamma^\mu \gamma^5$
and the definition in (\ref{wavefunctiondef}), one gets the pole
terms:
\begin{eqnarray}
\label{poletau} \Xi^{\tau}_{pole} &=&
  \frac{f_{\frac{1}{2}^+}f_{\frac{1}{2}^-} (-\tau)}{(2\bar{\Lambda}-\omega-i\epsilon)(2\bar{\Lambda}'-\omega'-i\epsilon)}
   (v-v')^\mu   ,\\
\label{poleb} \Xi^{\chi^b_0}_{pole} &=&
  \frac{f_{\frac{1}{2}^+}f_{\frac{1}{2}^-}}{(2\bar{\Lambda}-\omega-i\epsilon)(2\bar{\Lambda}'-\omega'-i\epsilon)}
  \frac{\chi^b_0}{\bar{\Lambda}} (v-v')^\mu  , \\
\label{polec} \Xi^{\chi^c_0}_{pole} &=&
  \frac{f_{\frac{1}{2}^+}f_{\frac{1}{2}^-}}{(2\bar{\Lambda}-\omega-i\epsilon)(2\bar{\Lambda}'-\omega'-i\epsilon)}
  \frac{\chi^c_0}{\bar{\Lambda}'} (v-v')^\mu .
\end{eqnarray}
$f_{\frac{1}{2}^+}$ and $f_{\frac{1}{2}^-}$ are the leptonic decay
constants of relevant mesons at leading order approximation:
\begin{eqnarray}
\langle 0|J_{0,+,1/2}|{D'_0}_{v'} \rangle &=& \sqrt{\bar{\Lambda}'} f_{+,1/2} , \\
\langle 0|J_{0,-,1/2}|B_{v} \rangle &=& \sqrt{\bar{\Lambda}}
f_{-,1/2}  .
\end{eqnarray}

In sum rule approach, the theoretical representation for the
correlation functions can be calculated from QCD or effective
theories in the deep Euclidean region, and in performing the
operator product expansion the nonperturbative effects are
incorporated via the inclusion of nonzero vacuum condensate
values. Formally the theoretical sides of the sum rules can be
written as
\begin{eqnarray}
\label{theo} \Xi^{\tau}_{theo}
(\Xi^{\chi^{b(c)}_0}_{theo})=\int d\nu
d\nu' \frac{ \rho^{\tau}_{pert} (\rho^{b(c)}_{pert})}
   {(\nu-\omega-i\epsilon)(\nu'-\omega'-i\epsilon)} +\Xi_{NP} +\mbox{subtraction
   terms}
\end{eqnarray}
with $\Xi_{NP}$ being the nonperturbative terms. QCD sum rules are
obtained by equating the phenomenological and theoretical
representations of the correlators. In doing this the perturbative
contribution above some threshold energy is assumed to simulate
the higher resonance contribution.
To suppress the higher resonance contribution and at the same time
enhance the importance of low dimension condensates, Borel
transformation
\begin{eqnarray}
 \label{eq:defBT}
   \hat{B}^{(\omega)}_{T}\equiv T  \lim_{n \to \infty, -\omega \to \infty}
   \frac{ \omega^n}{\Gamma(n)} (-\frac{d}{d\omega})^n
   \;\;\; \mbox{with} \;\; T=\frac{-\omega}{n} \;\; \mbox{fixed}
 \end{eqnarray}
should be performed to both sides of sum rules. Since there are
two variables $\omega$ and $\omega'$ for the correlation functions
(\ref{correlator1})-(\ref{correlator5}), we shall perform a double
Borel transformation
$\hat{B}^{(\omega)}_{t}\hat{B}^{(\omega')}_{t'}$, which then
introduces two Borel parameters $t$ and $t'$ in the sum rules. In
studying B decays into ground state charmed mesons, it is argued
\cite{neubprd46,bs,neurep} that the hadronic and perturbative
spectral densities can not be locally dual to each other, but the
quark-hadron duality is restored in the ``diagonal" variable
$\nu_{+}=\frac{\nu+\nu'}{2}$. Here we shall follow this
prescription. That is, we integrate the spectral densities over
the ``off-diagonal" variable $\nu_{-}=\frac{\nu-\nu'}{2}$, and
assume the quark-hadron duality in $\nu_{+}$ for the integrated
spectral densities. This can be represented as
\begin{eqnarray}
  \label{eq:IWphentheo}
   \tilde{\Xi}_{pole}=2\int^{s_0}_0 d\nu_{+} e^{-\nu_{+}/T}
      \tilde{\rho}_{pert}(\nu_{+})+\tilde{\Xi}_{NP},
  \end{eqnarray}
where the two Borel parameters are set equal, $t=t'=2T$.
$\tilde{\Xi}$ is obtained by applying double Borel operators to
$\Xi$, and
\begin{eqnarray}
    \tilde{\rho}_{pert}(\nu_{+})=\int^{\nu_{+}}_{-\nu_{+}} d\nu_{-} \rho_{pert}(\nu_{+},\nu_{-}).
  \end{eqnarray}

In the OPE we consider condensates with dimension no higher than
5, and the light quark mass and higher radiative corrections are
neglected. Then the Feynman diagrams presented in Fig.1 should be
calculated. The resulting sum rules turn out to be
\begin{eqnarray}
\label{srtau2}  f_{\frac{1}{2}^+} f_{\frac{1}{2}^-} \tau
e^{-(\bar{\Lambda}+\bar{\Lambda}')/T} &=& \frac{1}{8 \pi^2
(1+y)^2} \int^{s^\tau_0}_{0} d\nu_+ \nu_+^3
      e^{-\nu_+/T}
   -\frac{ 2 T \alpha_s }{3\pi}\langle \bar{q}q \rangle + \frac{1}{96 \pi^2 T}
    [ 6\pi^2 (y+2) \nonumber \\
    && - 4\pi  (y+1)\alpha_s ] i \langle \bar{q} \sigma_{\alpha\beta}
   F^{\alpha\beta} q \rangle +
    \frac{ (y-1)}{192 \pi (y+1)}\alpha_s \langle F^a_{\alpha\beta} F^{a\alpha\beta} \rangle , \\
\label{srb2}  f_{\frac{1}{2}^+} f_{\frac{1}{2}^-}
\frac{\chi^{b}_0}{\bar{\Lambda}}
e^{-(\bar\Lambda+\bar{\Lambda}')/T} & =&  - \frac{y+4}{16 \pi^2
(1+y)^3} \int^{s^{b}_0}_{0} d\nu_+    \nu_+^4     e^{-\nu_+/T}
  -\frac{5  T^2 \alpha_s}{3\pi (y+1)} \langle \bar{q}q\rangle  \nonumber\\
 && +\frac{ (y+2)T }{96 \pi (1+y)^2 } \alpha_s \langle F^a_{\alpha\beta} F^{a \alpha\beta} \rangle
     , \\
 \label{src2}
f_{\frac{1}{2}^+} f_{\frac{1}{2}^-}
\frac{\chi^{c}_0}{\bar{\Lambda}'}
e^{-(\bar\Lambda+\bar{\Lambda}')/T}
  &=& \frac{3(3y+2)}{16 \pi^2 (1+y)^3}
    \int^{s^{c}_0}_0 d\nu_+    \nu_+^4   e^{-\nu_+/T}
  - \frac{ (4y+3)T^2 \alpha_s}{3\pi (y+1)} \langle \bar{q}q \rangle \nonumber\\
  && - \frac{(y+8) T}{96 \pi (y+1)} \alpha_s \langle
F^a_{\alpha\beta} F^{a \alpha\beta} \rangle ,
\end{eqnarray}
where the threshold values should be determined by the principle
of minimal sensitivity in the numerical analysis of sum rules. The
condensates have the typical values:
 \begin{eqnarray}
   &&\langle \bar{q} q \rangle \approx -(0.23 \; \mbox{GeV})^3 ,  \nonumber\\
   &&i \langle \bar{q} \sigma_{\alpha \beta} F^{\alpha \beta} q \rangle \approx
   -m^2_0 \;
   \langle \bar{q} q \rangle \hspace{0.2cm} \mbox{ with } m^2_0=0.8 \; \mbox{GeV}^2  ,  \nonumber\\
   && \alpha_s  \langle F^a_{\alpha\beta} F^{a \alpha \beta} \rangle
   \approx 0.04 \;\mbox{GeV}^4 .
 \end{eqnarray}
Eqs.(\ref{currentf0}) and (\ref{currentz0}) are used as
interpolating currents in deriving the sum rules
(\ref{srtau2})-(\ref{src2}). We have also considered the current
(\ref{currentz0x}) but we find that using such current results in
zero contribution of the perturbative diagram (the first diagram
in Fig.1), which makes the resulting sum rule equation not
reliable. This has been noted in Ref.\cite{YBDai9807461} and we
just further checked it.

To derive the wave functions from these sum rules one needs to
know the leptonic decay constants $f_{\frac{1}{2}^-}$ and
$f_{\frac{1}{2}^+}$. They can also be evaluated in the same
framework through QCD sum rule approach. The sum rule for
$f_{\frac{1}{2}^-}$ has been analyzed by previous work
\cite{neubprd1076,WYWangconsi,neurep} and our result is
\cite{WYWangconsi}
\begin{eqnarray}
\label{sr1} f^2_{\frac{1}{2}^-}
e^{-2\bar{\Lambda}/T} &=& \frac{3}{16\pi^2} \int^{s^-_0}_{0} d\nu
  \nu^2  e^{-\nu/T}  -\frac{ 1 }{2} (1+\frac{4\alpha_s}{3\pi})\langle \bar{q}q \rangle
  -  \frac{ 1 }{8T^2}
  ( 1+\frac{4 \alpha_s}{\pi} ) i\langle \bar{q}\sigma_{\alpha\beta}F^{\alpha\beta} q \rangle \nonumber \\
  && -\frac{ 1 }{48 \pi
  T} \alpha_s \langle F^a_{\alpha\beta}F^{a \alpha\beta} \rangle ,
\end{eqnarray}
where the relation between $f_{\frac{1}{2}^-}$ and $F$ in
Ref.\cite{WYWangconsi} is $F=\sqrt{2}
 f_{\frac{1}{2}^-}$.

For $f_{\frac{1}{2}^+}$, we consider the two-point correlation
function
\begin{eqnarray}
\label{correlatorz} \Pi=i\int d^4x e^{ik\cdot x} \langle 0|T\{
J_{0,+,1/2}(x), J_{0,+,1/2}^\dagger (0)\} |0\rangle  .
\end{eqnarray}
Inserting a complete set of intermediate states and assuming the
quark-hadron duality, one has
\begin{eqnarray}
\frac{f_{\frac{1}{2}^+}^2}{2 \bar{\Lambda}'-2 v\cdot k -i\epsilon
}=\int^{s^+_0}_0 d \nu
\frac{\rho_{pert}(\nu)}{\nu-\omega-i\epsilon}+\Pi_{NP}+\mbox{subtraction
terms} .
\end{eqnarray}
After Borel transformation we get
\begin{eqnarray}
\label{sr3} f^2_{\frac{1}{2}^+}
e^{-2\bar{\Lambda}'/T}&=&\frac{3}{64\pi^2} \int^{s^+_0}_{0} d\nu
 \nu^4 e^{-\nu/T}  + ( \frac{3}{16}-\frac{\alpha_s}{32\pi} )i\langle
\bar{q}\sigma_{\alpha\beta} F^{\alpha\beta} q\rangle .
\end{eqnarray}
$f_{\frac{1}{2}^+}$ has also been studied in the usual HQET
\cite{YBDai9609436}. We note that the perturbation term in
Eq.(\ref{sr3}) is same as that in Ref.\cite{YBDai9609436}. Our
calculation includes contributions from all diagrams in Fig.2, and
the nonperturbative terms have some difference to that reference.

Sum rules in Eqs.(\ref{srtau2})-(\ref{src2}), (\ref{sr1}) and
(\ref{sr3}) constitute the main results that we will use to
discuss the $B\to D'_0(D'_1)\ell \bar{\nu} $ decays. The constants
$f_{\frac{1}{2}^-}$ and $f_{\frac{1}{2}^+}$ as well as the binding
energy $\bar{\Lambda}$ and $\bar{\Lambda}'$ can be estimated from
sum rules (\ref{sr1}) and (\ref{sr3}). And the wave functions
$\tau(y)$, $\chi^b_0(y)$ and $\chi^c_0(y)$ can be derived by
studying the ratios of Eqs.(\ref{srtau2})-(\ref{src2}) to
(\ref{sr1}) and (\ref{sr3}).

QCD higher order corrections are not included in our calculation.
They affect both the three-point and two-point correlation functions and deserve further study in future work.
As far as the determination of transition wave functions is concerned in
this paper, the effects of radiative corrections are expected to be partly cancelled in the ratios of
three-point to two-point correlators, and therefore not influence the
final results significantly. This has been proved to be true in the study of
Refs.\cite{PCola116005,neub4063,EBagan249}.
In those references the two-loop corrections to the Isgur-Wise functions are found to be small and well under control for the B decays into both ground state \cite{neub4063,EBagan249} and excited state \cite{PCola116005}
charmed mesons, although the corrections to decay constants
are sizable.

\section{Numerical Results}\label{analysis}

We get from Eq.(\ref{sr1}) the appropriate binding energy and
decay constant as \cite{WYWangconsi}
\begin{eqnarray}
&& \bar{\Lambda}_{\frac{1}{2}^-}=0.53 \pm 0.08 \mbox{GeV}, \nonumber \\
&& f_{\frac{1}{2}^-}= 0.21\pm 0.05 \mbox{GeV}^{3/2} .
\end{eqnarray}

For $f_{\frac{1}{2}^+}$ one should study Eq.(\ref{sr3}).
$\bar{\Lambda}_{\frac{1}{2}^+}$ and $f_{\frac{1}{2}^+}$ as
functions of the Borel parameter $T$ is presented in Fig.3.
$\bar{\Lambda}_{\frac{1}{2}^+}$ and $f_{\frac{1}{2}^+}$ have
acceptable stability when setting the threshold
$s^+_0=2.6-3.0\mbox{GeV}$. The curves in Fig.3 become rather
stable when $T>1\mbox{GeV}$. However, the criterion of sum rule
analysis is that both contributions from the higher resonances and
from the higher order power corrections in OPE should not be very
large, say not much higher than 30\%. According to this criterion
the proper window for Eq.(\ref{sr3}) is $0.6\mbox{GeV} < T <
0.8\mbox{GeV}$. As a result we get
\begin{eqnarray}
\label{lambaz}
&& \bar{\Lambda}_{\frac{1}{2}^+}=0.81 \pm 0.12 \mbox{GeV}, \nonumber \\
&& f_{\frac{1}{2}^+}= 0.30 \pm 0.05 \mbox{GeV}^{5/2},
\end{eqnarray}
where the central values are obtained using $s^+_0=2.8$GeV and
$T=0.7$GeV, and the errors are attributed to the variation of the
threshold and Borel parameter.

Now we can study the sum rules for $B\to D'_0(D'_1)\ell\nu $ wave
functions. The leading function $\tau$ depends on the recoil
variable $y$ and can be estimated from the sum rule
(\ref{srtau2}). Fig.4 displays $\tau$ as a function of the Borel
parameter at the fixed point $y=1$. Applying the sum rule
criterion the appropriate region for analyzing the stability is
$0.8\mbox{GeV} <T<1.0 \mbox{GeV} $. As can be seen in the figure,
$2.7\mbox{GeV}<s^\tau_0<3.3\mbox{GeV}$ is favorable. Therefore we
get
\begin{eqnarray}
\tau(1) = 0.57 \pm 0.06,
\end{eqnarray}
where the central value corresponds to $s^\tau_0 = 3.0 \mbox{GeV}$
and $T=0.9 \mbox{GeV}$.

Following the same procedure the subleading order wave functions
$\chi^{b}_0$ and $\chi^{c}_0$ can be derived from Eqs.(\ref{srb2})
and (\ref{src2}). The results at zero recoil are shown in Fig.5
and Fig.6, respectively. In the appropriate window $\chi^{b}_0(1)$
is not sensitive to the Borel parameter when $s^b_0 \sim
2.1\mbox{GeV}$, while $\chi^{c}_0(1)$ becomes stable around a
smaller threshold value $s^c_0 \sim 1.2 \mbox{GeV}$. Setting $T
\sim 0.9 \mbox{GeV} $ we then obtain the following zero recoil
values for the $1/m_Q$ order wave functions
\begin{eqnarray}
\label{functionvalueb}
-\frac{\chi^{b}_0(1)}{\bar{\Lambda}}&=& 0.27 \pm 0.12  \mbox{GeV}, \\
\label{functionvaluec} -\frac{\chi^{c}_0(1)}{\bar{\Lambda}'}&=&
-0.20  \pm 0.12 \mbox{GeV},
\end{eqnarray}
where the errors mainly arise from the thresholds. So $\chi^b_0$
and $\chi^c_0$ have opposite signs. As $m_b > m_c$, $\chi^c_0$ may
yield a relatively larger contribution to the $B\to D'_0 (D'_1)$
form factors. $\chi^b_0$ can only weakly counteract the
contribution of $\chi^c_0$, which makes $\tilde{\tau}$ in
Eq.(\ref{taucor}) suppressed with respect to $\tau$.

If one fix the values of the thresholds and the parameter T,
$\tau$ and $\chi^{b(c)}_0$ as functions of the recoil value can be
evaluated from the sum rule equations. The results are shown in
Fig.7, where $T=0.9\mbox{GeV}$ is used. We find these functions
can be expanded near $y=1$ as
\begin{eqnarray}
\label{tauy}
\tau(y)&=&\tau(1) [ 1-0.56 (y-1)+0.35 (y-1)^2  ] ,\\
\label{chi0by} \chi^b_0(y)&=&\chi^b_0 (1)[ 1-1.45 (y-1)+0.98
(y-1)^2 ],
\\
\label{chi0cy} \chi^c_0(y)&=&\chi^c_0(1) [ 1-0.52 (y-1)+0.28
(y-1)^2 ].
\end{eqnarray}

When $\chi^{b(c)}_{1(2)}$ are neglected, the form factors in
(\ref{formfactorHQE}) can be simply written as
\begin{eqnarray}
\label{gsim} g_+ &=& 0 ,\hspace{2cm} g_- = \hat{\tau}_{D'_0}
,\hspace{2cm}
g_{V_1} = (y-1) \hat{\tau}_{D'_1}, \nonumber \\
g_{V_2} &=& 0, \hspace{1.9cm} g_{V_3} =-\hat{\tau}_{D'_1},
\hspace{1.8cm}
 g_{A} =  \hat{\tau}_{D'_1}
\end{eqnarray}
with the definition
\begin{eqnarray}
\label{tauhatD0} \hat{\tau}_{D'_0} &=&
\tilde{\tau}+\frac{\tau}{2m_b\bar{\Lambda} }
  (\kappa_1(1)+3\kappa_2(1))+\frac{\tau}{2m_c \bar{\Lambda}'}
  (\kappa'_1(1)+3\kappa'_2(1)) ,  \\
\label{tauhatD1} \hat{\tau}_{D'_1} &=&
\tilde{\tau}+\frac{\tau}{2m_b\bar{\Lambda} }
  (\kappa_1(1)+3\kappa_2(1))+\frac{\tau}{2m_c \bar{\Lambda}'}
  (\kappa'_1(1)-\kappa'_2(1))   .
\end{eqnarray}
Consequently the differential decay rates turn into
\begin{eqnarray}
\frac{d\Gamma(B\to D'_0 \ell \bar\nu)}{dy}&=& \frac{G^2_F
|V_{cb}|^2 m^5_B }{48 \pi^3} {r'_0}^3 (1-r'_0)^2
   (y^2-1)^{3/2}   \hat{\tau}^2_{D'_0} \; , \nonumber \\
\frac{d\Gamma(B\to D'_1 \ell \bar\nu)}{dy}&=& \frac{G^2_F
|V_{cb}|^2 m^5_B }{48 \pi^3} {r'_1}^3
   (y^2-1)^{1/2} [ (1+{r'_1}^2)(5y^2 - 6y+1) \nonumber \\
   &&-r'_1 (8y^3-10y^2+4y-2) ]
    \hat{\tau}^2_{D'_1} .
\end{eqnarray}

$\kappa^{(\prime)}_i(1) (i=1,2)$ are parameters related to meson
masses. Taking $m_b=4.67 \pm 0.05\mbox{GeV}$, $m_c=1.35 \pm 0.05\mbox{GeV}$,
$m_B=5.279\mbox{GeV}$, $m_{B^*}=5.325\mbox{GeV}$ and the averaged masses for the $\frac{1}{2}^+$
doublet, $m_{D'_0}=2.351 \pm 0.027\mbox{GeV}$ and $m_{D'_1}=2.438 \pm 0.030 \mbox{GeV}$
\cite{PCola0407137}, we get from
Eqs.(\ref{binding1})-(\ref{binding4})
\begin{eqnarray}
\label{kappavalue} \kappa_1(1) &=& -0.53 \pm 0.23 \mbox{GeV}^2 ,
\hspace{2cm}
\kappa_2(1)=0.05 \pm 0.01 \mbox{GeV}^2 , \nonumber \\
\kappa'_1(1) &=& -0.35 \pm 0.17 \mbox{GeV}^2 , \hspace{2cm}
\kappa'_2(1)=0.03 \pm 0.01 \mbox{GeV}^2  ,
\end{eqnarray}
where we include uncertainties from the binding energies and the quark and meson masses.
The values of $\kappa_1(1)$ and $\kappa_2(1)$ in (\ref{kappavalue}) are consistent with the
results of Ref.\cite{WYWangconsi} in which these two parameters are evaluated
through QCD sum rule equations that are independent of the heavy quark and meson masses.
$\kappa^{(\prime)}_2(1)$ characterize the mass splittings of two
mesons belonging to a $j^P_l$ doublet, and the absolute values of
them are much smaller than those of $\kappa^{(\prime)}_1(1)$.
Therefore one gets from Eqs.(\ref{tauhatD0}) and (\ref{tauhatD1})
$\hat{\tau}_{D'_0} \approx \hat{\tau}_{D'_1}$. It is also clear
from (\ref{kappavalue}) that $\hat{\tau}_{D'_0}$ and $
\hat{\tau}_{D'_1}$ may be further suppressed by
$\kappa^{(\prime)}_1$ with respect to $\tilde{\tau}$ and $\tau$.

With the obtained values of wave functions, we get the
decay rates and branching ratios in Table 1. In the calculation
the B meson life time $\tau(B)=1.542 \; \mbox{ps}$ and
$|V_{cb}|=0.041$ are used, and the uncertainties arise from the quark
and meson masses as well as the variation of
the thresholds and the Borel parameter. It is shown that the branching ratios
in the $m_Q\to \infty$ limit can be about $1\times 10^{-3}$.
However $\chi^b_0$ and $\chi^c_0$ may lead to nearly 20\%
suppression. When the contribution of $\kappa^{(\prime)}_i(1)$ is
included, the branching ratios can even be significantly reduced.
We note that the masses of $\frac{1}{2}^+$ charmed doublets have
not been determined precisely. Consequently the values of
$\kappa'_i(1)$ may suffer from larger
uncertainty. So do the data in the last column of Table 1.
Nevertheless the substantial suppression effect of the $1/m_Q$
contribution to the decay rates is evident.

\begin{center}
\begin{tabular}{c|c|c|c|c}
\hline \hline
 \multirow{2}{*}{} & \multirow{2}{*}{\hspace{1cm}} & \multirow{2}{*}{$\;\; m_Q \to \infty \;\;\mbox{limit}\;\; $}
   & with $1/m_Q$ correction &   with $1/m_Q$ correction  \\
&  &  & from $\chi^Q_0$ & from $\chi^Q_0$ and $\kappa^{(\prime)}_i$      \\
\hline
\multirow{2}{*}{$B \to D'_0 \ell \bar{\nu} $} & $\Gamma$ & $ 4.14\pm 1.20 $  & $3.45\pm 1.09 $
   & $2.13 \pm 0.69 $ \\
\cline{2-5}
 & Br  & $ 0.97 \pm 0.28 $ & $ 0.81 \pm 0.26 $ & $ 0.50 \pm 0.16 $ \\
\hline
\multirow{2}{*}{$B \to D'_1 \ell \bar{\nu} $} & $\Gamma$ & $4.65 \pm 1.32$ & $3.89 \pm 1.20$
   & $2.05 \pm 0.66 $ \\
\cline{2-5}
 & Br & $1.09 \pm 0.31$  & $0.92 \pm 0.28$ & $0.48 \pm 0.16$ \\
\hline  \hline
\end{tabular}
\end{center}

 \vspace{0cm}
\centerline{
\parbox{12cm}{
\small \baselineskip=1.0pt Table 1. Rates $\Gamma$ (in units of
$|V_{cb}/0.041|^2\times 10^{-16}$ GeV) and branching ratios (in
$10^{-3}$) of $B\to D'_0(D'_1)\ell \bar{\nu}$ decays in the
$m_Q\to \infty$ limit as well as when taking account of $1/m_Q$
order corrections from $\chi^Q_0$ and $\kappa^{(\prime)}_i(1)$. }}

\section{Summary}\label{summary}

Semileptonic B decays into $j_l^P=\frac{1}{2}^+$ doublet excited
charmed mesons are studied in the framework of heavy quark
effective field theory with inclusion of the heavy quark-antiquark
coupling effects in the finite mass corrections. We present the
heavy quark expansion for $B\to D'_0(D'_1)$ transition matrix
elements. At the leading order of HQE all form factors reduce to
the leading Isgur-Wise function $\tau$, and at $1/m_Q$ order there
are six wave functions $\chi^{b(c)}_i\; (i=0,1,2)$. Among them
$\chi^{b(c)}_0$ characterize the contribution from the kinetic
energy operators and are expected to be much larger than
$\chi^{b(c)}_{1(2)}$ that are related to the chromomagnetic
operators. Beside these functions characterizing transitions
between $\frac{1}{2}^-$ and $\frac{1}{2}^+$ doublets, we present
also the functions $\kappa'_i(i=1,5)$ for transitions between
$\frac{1}{2}^+$ mesons. The zero recoil values of $\kappa'_1$ and
$\kappa'_2$ are extracted from the excited meson masses.

QCD sum rule method is applied to evaluate the functions $\tau$,
$\chi^b_0$ and $\chi^c_0$, and the decay rates are predicted. The
Isgur-Wise function $\tau$ gives the branching ratios of the
magnitude $10^{-3}$. However, the ratios are suppressed rather
than enhanced by the $1/m_Q$ corrections. Though we have not
calculated all $1/m_Q$ order wave functions, our results imply
that the finite mass corrections would not likely change the
dominance of the semileptonic B decay rates to $\frac{3}{2}^+$
states over the rates to $\frac{1}{2}^+$ states. As a result, the
production of $\frac{1}{2}^-(D, D^*)$, $\frac{3}{2}^+(D_1, D^*_2)$
and $\frac{1}{2}^+(D'_0, D'_1)$ charmed mesons do not saturate the
total semileptonic decay rate $\Gamma_{SL}(B)$, and the
configuration for the ``missing rate" remains an interesting
question. This is in agreement with the conclusion of the early work \cite{AKLeib}
that adopted some model dependent assumptions.

As for the recent report of BELLE \cite{BELLE091503} there is no
indication of a broad $D'_1$ in the $B\to D^* \pi \ell \bar{\nu}$
channel, but the measurements indicate for $B\to D'_0 \ell
\bar{\nu}$ a large rate of similar magnitude to the
$\frac{3}{2}^+$ production rates. If that result is confirmed, the
framework to predict $\frac{1}{2}^+$ production should be improved
to connect the theories and measurements. Anyway, according to the
calculations those rates might be in the reach of B facilities.
Measurements on such processes will test the theories and shed
light on the nature of excited states.

\acknowledgments

The author would like to thank Prof. Y. L. Wu for stimulating
discussions. This work was supported in part by the National
Science Foundation of China (NSFC) under the grant No. 10805005;
and by the Project of Knowledge Innovation Program (PKIP) of Chinese
Academy of Sciences, Grant No. KJCX2.YW.W10.

\newpage

\small

\hspace{1cm} \epsfxsize=15cm \epsfysize=20cm \epsffile{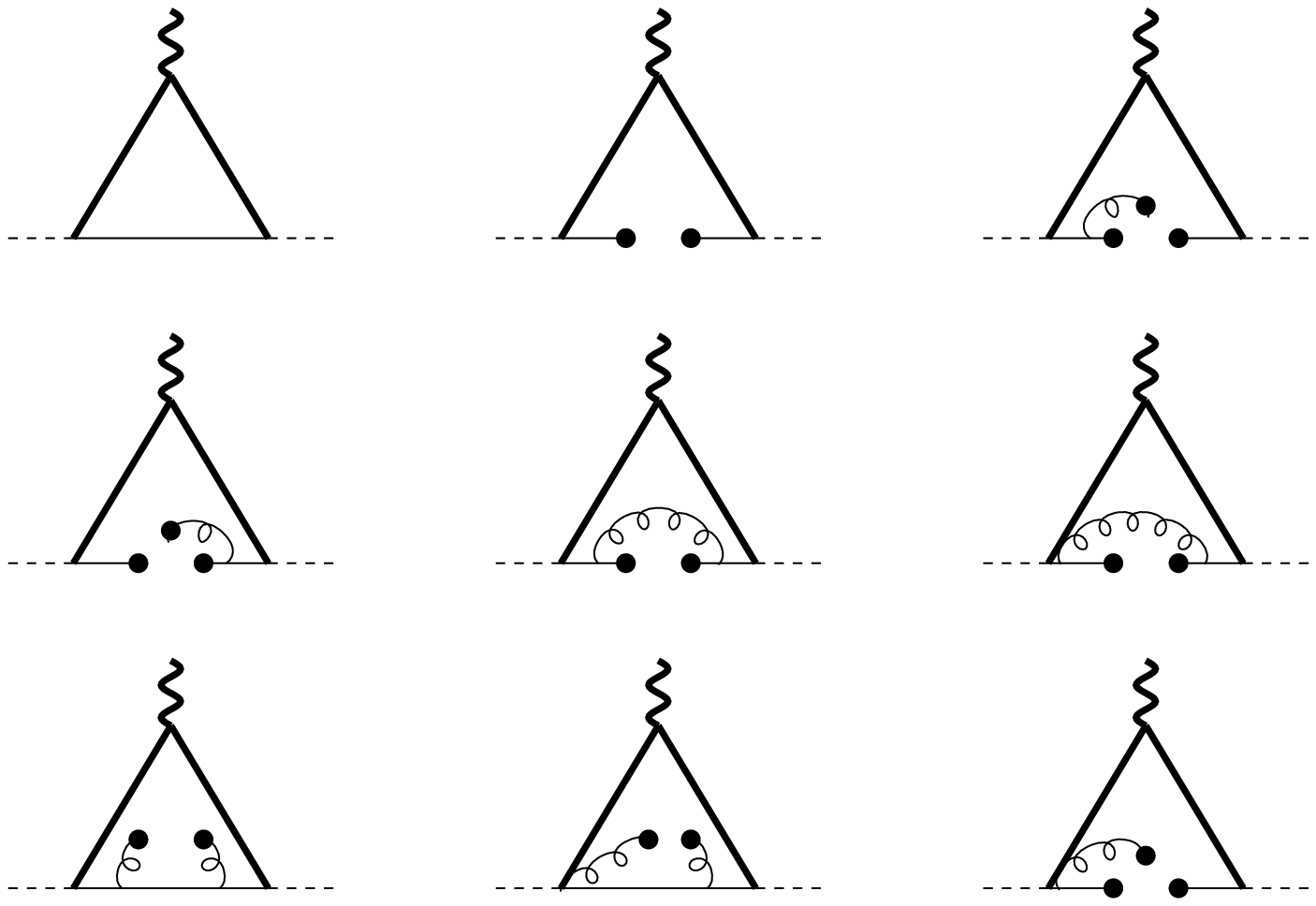}

\vspace{-12cm}
\centerline{
\parbox{12cm}{
\small \baselineskip=1.0pt Fig.1. Feynman diagrams contributing to
the sum rules for $\tau$, $\chi^b_0$ and $\chi^c_0$. The thick
lines represent heavy quarks; the light lines are light quarks;
the curves are gluon fields; the black dots represent condensates;
and the external lines represent the currents in
Eqs.(\ref{correlator1})-(\ref{correlator5}). }}

\hspace{1cm} \epsfxsize=15cm \epsfysize=20cm \epsffile{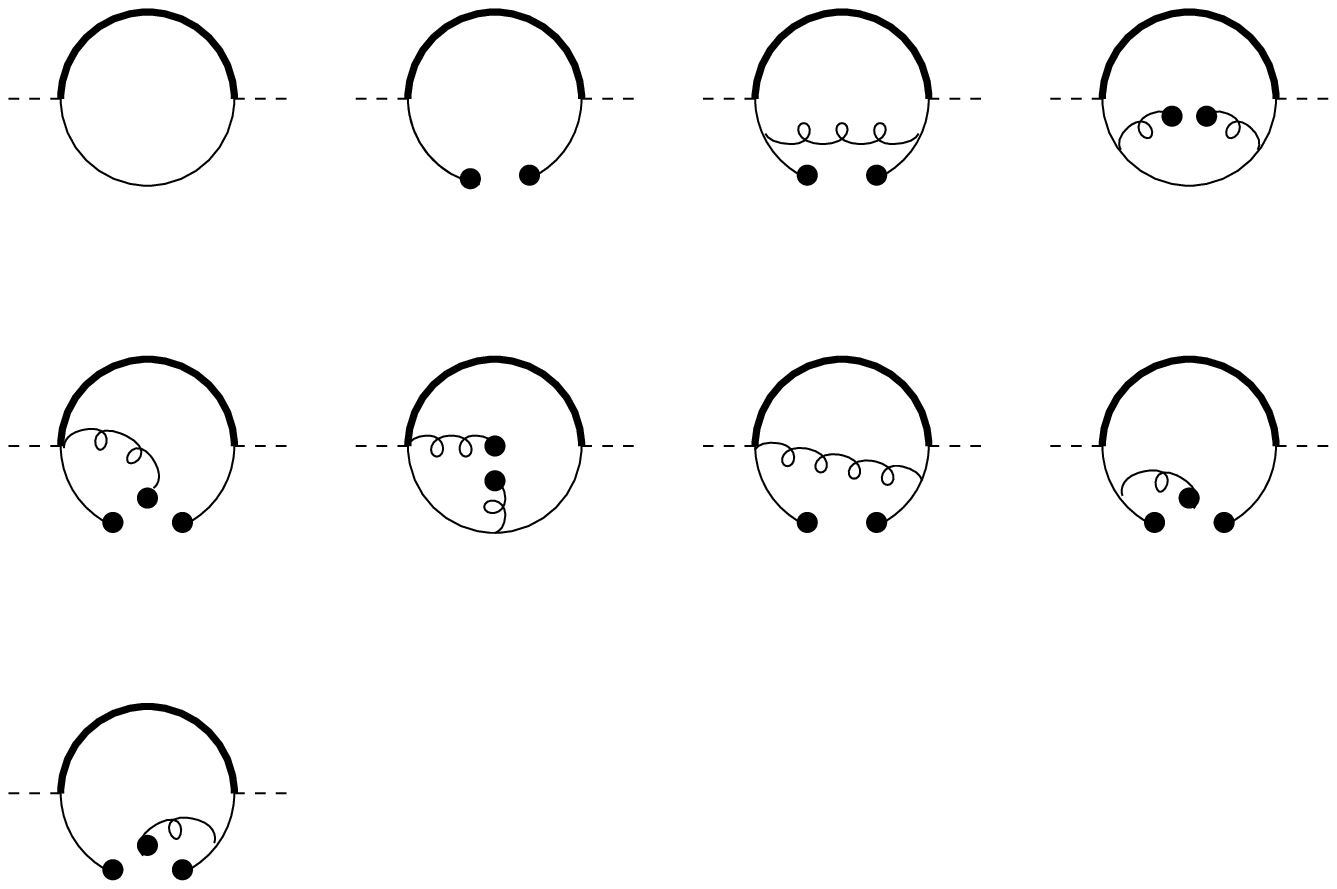}

\vspace{-13cm} \centerline{
\parbox{12cm}{Fig.2. Feynman diagrams contributing to the sum rule for
$f_{\frac{1}{2}^+}$. The external dashed lines represent the
interpolating currents in Eq.(\ref{correlatorz}). \small
\baselineskip=1.0pt  }}

\newcommand{\PIC}[2]
{
\begin{center}
\begin{picture}(300,220)(0,0)
\put(20,25){ \epsfxsize=8cm \epsfysize=6cm \epsffile{#1} }
\put(150,10){\makebox(0,0){#2}}
\end{picture}
\end{center}
}

\small \mbox{}

\PIC{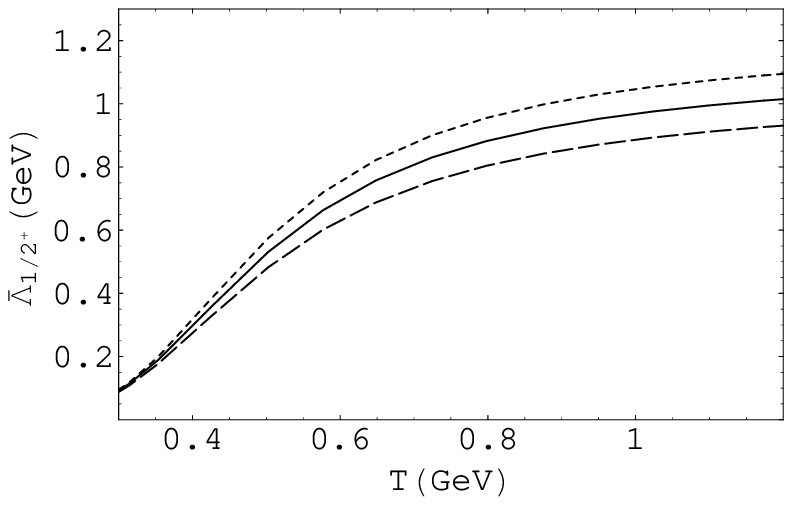}{(a)}

\PIC{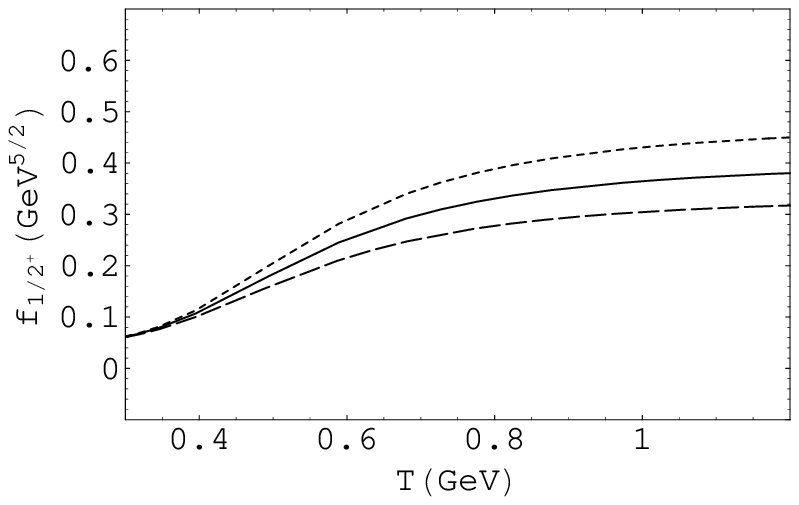}{(b)}

\vspace{0cm} \centerline{
\parbox{12cm}{
\small \baselineskip=1.0pt Fig.3. $\bar{\Lambda}_{\frac{1}{2}^+}$
(figure (a)) and $f_{\frac{1}{2}^+}$ (figure (b)) as functions of
Borel parameter T. The dashed, solid and dotted curves correspond
to $s^+_0$=2.6, 2.8 and 3.0 GeV, respectively. }}

\vspace{1cm}

\small \mbox{}

\PIC{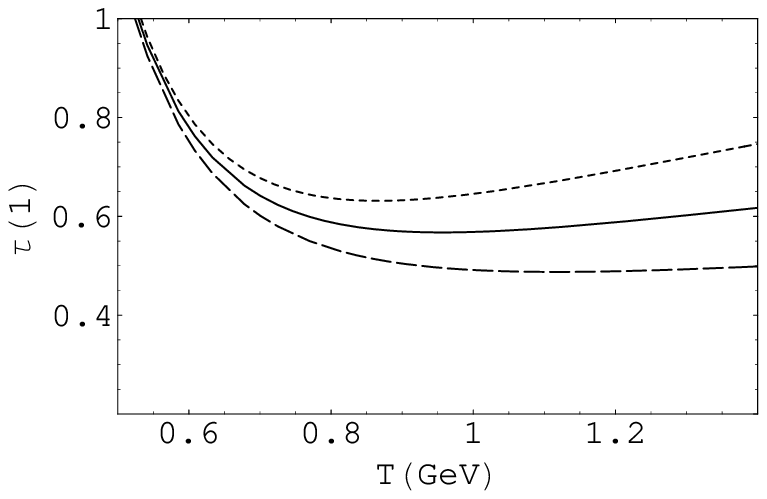}{}

\vspace{-1cm} \centerline{
\parbox{12cm}{
\small \baselineskip=1.0pt Fig.4. $\tau(1)$ as a function of the
Borel parameter T. The dashed, solid and dotted curves correspond
to $s^\tau_0$=2.7, 3.0 and 3.3 GeV, respectively. }}

\vspace{1cm}

\small \mbox{}

\PIC{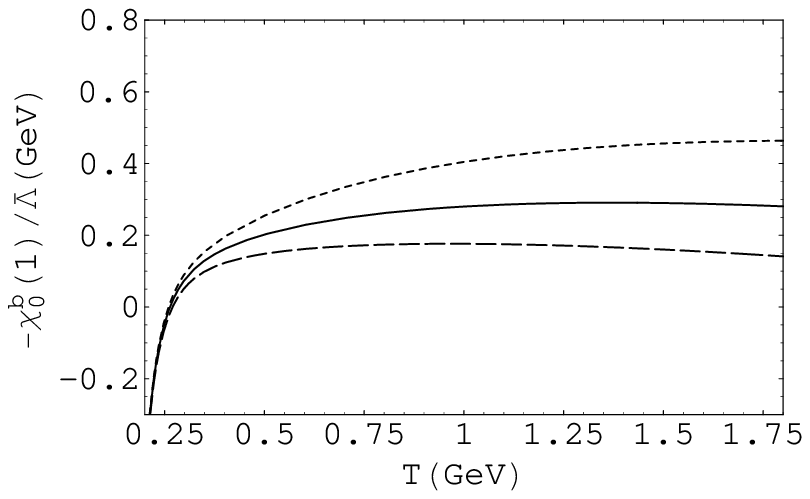}{}

\vspace{-1cm} \centerline{
\parbox{12cm}{
\small \baselineskip=1.0pt Fig.5. $-\chi^b_0(1)/\bar{\Lambda}$ as
a function of the Borel parameter T. The dashed, solid and dotted
curves correspond to $s^b_0$=1.9, 2.1 and 2.3 GeV, respectively.
 }}

\vspace{1cm}

\small \mbox{}

\PIC{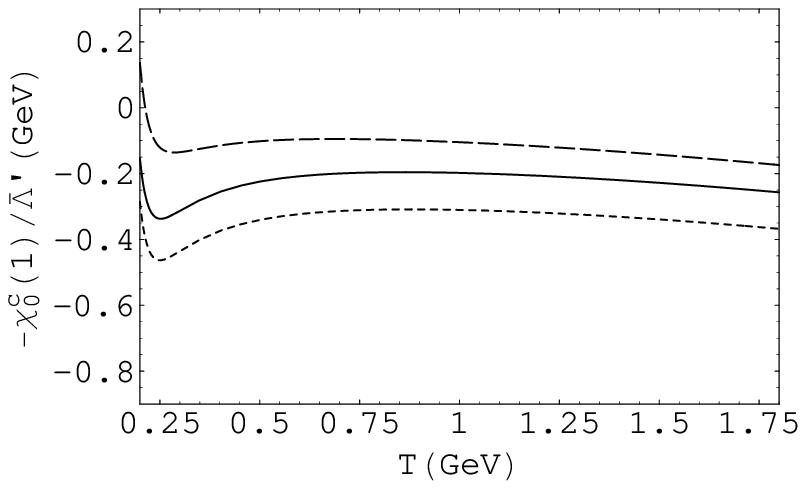}{}

\vspace{-1cm} \centerline{
\parbox{12cm}{
\small \baselineskip=1.0pt Fig.6. $-\chi^c_0(1)/\bar{\Lambda}'$ as
a function of the Borel parameter T. The dashed, solid and dotted
curves correspond to $s^c_0$=0.9, 1.2 and 1.4 GeV, respectively.
}}

\vspace{1cm}

\small \mbox{}

\PIC{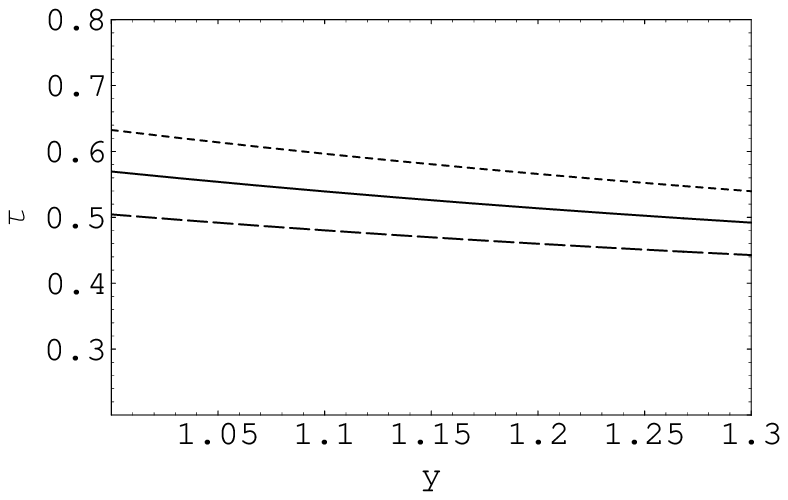}{(a)}

\PIC{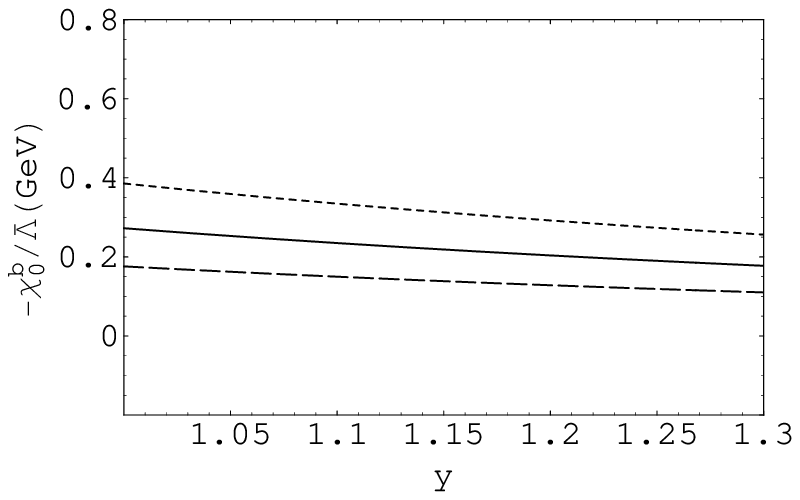}{(b)}

\PIC{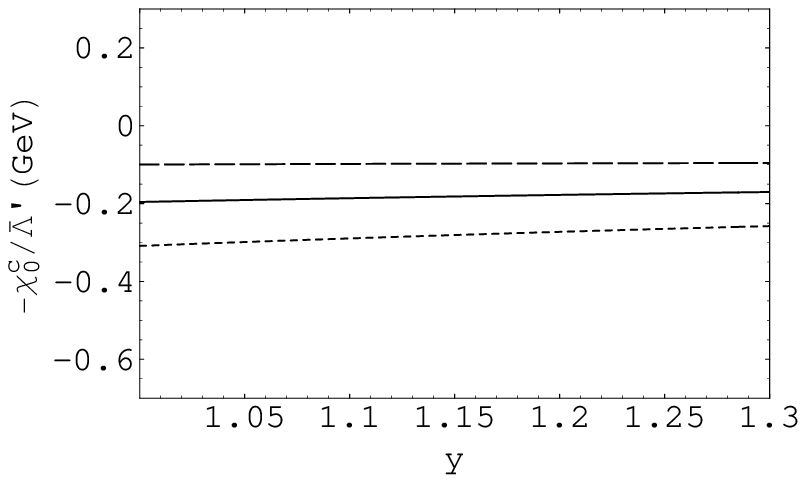}{(c)}

\vspace{0cm} \centerline{
\parbox{12cm}{
\small \baselineskip=1.0pt Fig.7. $\tau$,
$-\chi^b_0/\bar{\Lambda}$ and $-\chi^c_0/\bar{\Lambda}'$ as
functions of the variable y. The dashed, solid and dotted curves
correspond to $s^\tau_0$=2.7, 3.0 and 3.3 GeV in (a); $s^b_0
$=1.9, 2.1 and 2.3 GeV in (b); and $s^c_0$=0.9, 1.2 and 1.4 GeV in
(c). In the evaluation the Borel parameter T is set to 0.9 GeV.}}

\end{document}